\documentclass{skads2009}
\usepackage{graphicx}
\begin{document}
   \title{Mapping the SKA Simulated Skies with the $S^3$-Tools
	}

   \author{F. Levrier\inst{1}
          \and
          R.J. Wilman\inst{2}
          \and
          D. Obreschkow\inst{3}
          \and
          H.-R. Kl\"ockner\inst{3}
          \and
          I. Heywood\inst{3}
          \and
          S. Rawlings\inst{3}
          }

   \institute{LERMA/LRA - UMR 8112 - Ecole Normale Sup\'erieure, 24 rue Lhomond, 75231 Paris CEDEX 05, France
\and
Centre for Astrophysics \& Supercomputing, Swinburne University of Technology, Victoria 3122, Australia
   \and
     Oxford Astrophysics, Denys Wilkinson Building, Keble Road, Oxford, OX1 3RH, United Kingdom}

   \abstract{
The $S^3$-Tools are a set of Python-based routines and interfaces whose purpose is to provide user-friendly access to the SKA Simulated Skies ($S^3$) set of simulations, an effort led by the University of Oxford in the framework of the European Union's SKADS program ({\tt http://www.skads-eu.org}). The databases built from the $S^3$ simulations are hosted by the Oxford e-Research Center (OeRC), and can be accessed through a web portal at {\tt http://s-cubed.physics.ox.ac.uk}. This paper focuses on the practical steps involved to make radio images from the $S^3$-SEX and $S^3$-SAX simulations using the $S^3$-Map tool and should be taken as a broad overview. For a more complete description, the interested reader should look up the user's guide. The output images can then be used as input to instrument simulators, e.g. to assess technical designs and observational strategies for the SKA and SKA pathfinders.
}
   \maketitle
%
%
\section{Introduction}

The SKA Simulated Skies project ($S^3$) aims at building mock radio skies suitable for planning science with the SKA and the many pathfinder experiments. The project comprises a number of simulations, of which essentially two are relevant for this paper:

$\bullet$ $S^3$-SEX ({\bf S}emi-{\bf E}mpirical e{\bf X}tragalactic) : a large-scale simulation of the extragalactic radio continuum sky covering a sky area of $20^{\circ} \times 20^{\circ}$, out to redshift $z=20$, and down to 10 nJy. It also includes {\sc Hi} gas masses for star-forming galaxies. For a complete description, see \cite{wilman08}.

$\bullet$ $S^3$-SAX ({\bf S}emi-{\bf A}nalytical e{\bf X}tragalactic) : a smaller-scale simulation of the extragalactic {\sc Hi} and CO line emissions, derived from the Millenium simulation (\cite{springel05}), which comes in two flavours. $S^3$-SAX-Sky is a skyfield simulation giving the apparent properties of galaxies, with a field of view that depends on the maximum redshift, while $S^3$-SAX-Box is a simulation of a cubic volume giving their intrinsic properties. In all of this paper, $S^3$-SAX shall be understood as meaning $S^3$-SAX-Sky. For a complete description, see \cite{obreschkow09}.



\section{Querying the $S^3$ databases}

\subsection{Database structure}
The databases can be queried via SQL forms on the dedicated $S^3$-SEX and $S^3$-SAX sections of the $S^3$ website hosted at {\tt http://s-cubed.physics.ox.ac.uk}. The queries should be written in the SQL format, and query examples are given.

The $S^3$-SEX and $S^3$-SAX databases have similar structures, and consist of respectively three and two tables.

$\bullet$ For $S^3$-SEX : a {\sf Clusters} table holding cluster properties, a {\sf Galaxies} table holding galaxy properties (possibly including the cluster index to which they belong) and a {\sl Components} table, which lists the properties of the several components that may make up a single galaxy, such as cores and radio lobes.

$\bullet$ For $S^3$-SAX : a {\sf galaxies\_line} table containing the apparent position and emission line properties of the galaxies and a {\sf galaxies\_delu} table containing the intrinsic properties of the \cite{delucia06} catalog.


For more information on the structure of the database and the attributes held in the various tables, please refer to the appropriate sections of the $S^3$ website. It is advisable to do so before reading on, as in the following we shall mention some of these attributes explicitly.

\subsection{Mandatory attributes}
\label{sec:mandatoryitems}
To make maps from these simulations, it is mandatory that some properties - listed below - be included in the query, as the mapping algorithms require them :

$\bullet$ For $S^3$-SEX : 

\noindent
In the {\sf Galaxies} table : {\sf galaxy}, {\sf redshift} or {\sf modified\_redshift}, {\sf distance}, {\sf sftype}, {\sf agntype} (these latter two are necessary if one wishes to make separate images for the different types of sources), {\sf m\_hi} (if {\sc Hi} mapping is to be performed).
\\
In the {\sf Components} table : {\sf galaxy}, {\sf right\_ascension}, {\sf declination}, {\sf position\_angle}, {\sf major\_axis}, {\sf minor\_axis}, {\sf i\_151}, {\sf i\_610}, {\sf i\_1400}, {\sf i\_4860}, {\sf i\_14000}.

$\bullet$ For $S^3$-SAX : 

\noindent
In the {\sf Galaxies} table only : {\sf ra}, {\sf decl}, {\sf zapparent}, {\sf diskpositionangle}, {\sf diskinclination}, {\sf rmolc}, {\sf distance}. To these, one should add {\sf hiintflux}, {\sf himajoraxis\_10max}, {\sf hiwidth50} and {\sf hiwidth20} for mapping the {\sc Hi} line, as well as {\sf cointflux\_$J$}, {\sf h2majoraxis\_10max}, {\sf cowidth50} and {\sf cowidth20} for mapping the CO($J\to J-1$) line. When using the Kapteyn {\sc Hi} templates (see below), {\sf hubbletype} should also be retrieved.

\subsection{Structure of query result files}

Query results are saved in a gzipped tarball {\tt id.tar.gz}, where {\tt id} is a hash given by the server at submission time. The tarball contains two plain text files : {\tt id.sql}, in which the original SQL-formatted query is recalled, and {\tt id.result}, which contains the comma-separated query results proper.

\section{Installing the $S^3$-Tools}

\subsection{Prerequisites and installation}
The $S^3$-Tools are currently hosted on an external website at {\tt http://www.lra.ens.fr/$\sim$levrier/Recherche/S3/}, although they should find a permanent home on the $S^3$ webserver in the near future.

To install the $S^3$-Tools, you should refer to that website or to the README file included in the distribution. In particular, the user should be aware that the $S^3$-Tools require a number of python libraries, namely the scipy, numpy, os, math and pyfits packages. For Mac users, these are all included in the SciSoft package (at least from version 2008.3.1 onwards) for OS X.

Provided these packages are functioning properly on your system, installation is a simple matter of untarring the {\tt S3Tools.tgz} archive and setting a few paths in the {\tt Config\_Path.py} file in the {\tt Config/} subdirectory. In the following, we shall assume that installation is done in a directory named {\tt S3Tools} under your home directory. Configuration files should therefore be in {\tt $\sim$/S3Tools/Config/} and the main routines in {\tt $\sim$/S3Tools/Routines/}.

\subsection{Templates}
\label{sec:templates}
To make line maps from $S^3$-SAX, {\sc Hi} and CO templates are required. These were made by D. Obreschkow and are available on the same webpage. Be warned that the archive weighs over 2~GB. A different set of templates for {\sc Hi} has been produced by R. Boomsma from Kapteyn Institute and are also readily available to use.

\subsection{Global Sky Model}
\label{sec:gsm}

The Global Sky Model (GSM) - also dubbed $S^3$-GAL on the $S^3$ website - is a model of the Galactic foregrounds from 10 MHz to 100 GHz by \cite{deoliveiracosta08}. It is possible to include these foregrounds in spectral cubes built from the SKA Simulated Skies with $S^3$-Map. To be able to use this option, the GSM data and routines should be installed from the GSM website {\tt http://space.mit.edu/home/angelica/gsm/}, and the installation directory specified via the {\tt GSM\_DIR} variable in {\tt Config\_Path.py}.

\section{Making maps and spectral data cubes}

\subsection{The $S^3$-Map GUI and underlying scripts}

\begin{figure}
\centering
\includegraphics[width=7cm]{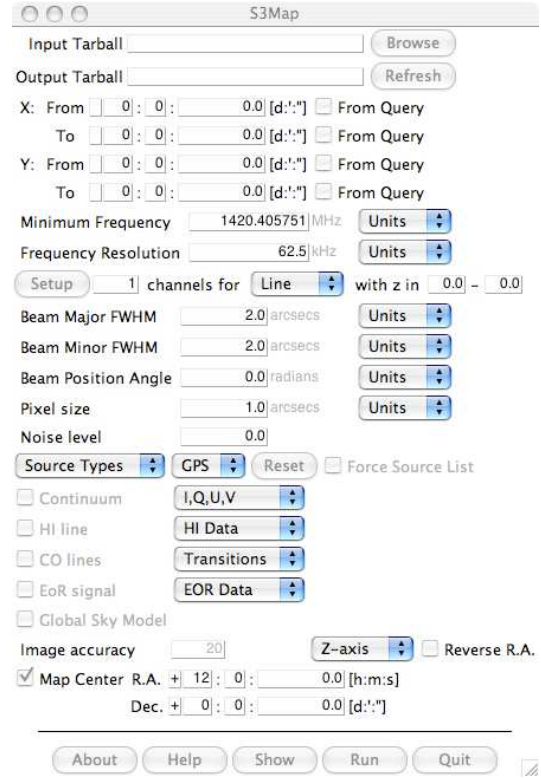}
\caption{The $S^3$-Map graphical user interface}
\label{fig:S3Map}
\end{figure}

Making maps from query results is done via $S^3$-Map, a graphical user interface (GUI) written in python, which is run by moving to the {\tt Routines/} subdirectory and typing {\tt python S3Map.py} at the command line. A window similar to the one presented in Fig.~\ref{fig:S3Map} should appear after a few moments. It is through this interface that users specify which maps to make. To be more specific, the state of the GUI is passed as arguments to a python script command, either {\tt SEX\_map\_script.py} or {\tt SAX\_map\_script.py}. It is this command that is in charge of actually building the maps using lower-level routines, and it may be printed out on the screen by clicking on the {\sf Show} button. The user can then copy-paste this command into a terminal window and perform the same mapping non-interactively. This is a useful trick to know for building batch jobs or debugging purposes. For a lengthy description of the syntax used for the arguments, please refer to the $S^3$-Tools user's guide.

\subsection{Common workings of the $S^3$-Map GUI}

Click on the {\sf Browse} button to select a tarball containing results from a database query. This prompts $S^3$-Map to check which database was queried, so that some of the buttons, menus, entries and options become either active or inactive, depending on their relevance. $S^3$-Map also fills in the mapping area fields with extremal values for right ascension ({\sf X}), declination ({\sf Y}) and redshift ({\sf z}) found in the query results. The {\sf From Query} checkboxes are then set, but the user may deactivate them and manually specify  a different mapping area.

Regarding frequency configuration, this is done via three entry fields : {\sf Minimum Frequency} (central frequency of the first channel), {\sf Frequency Resolution} (full channel width) and the number of channels. For line mapping, a convenient\footnote{"That {\sf Setup} button is a stroke of genius" (I. Heywood {\it dixit}).} {\sf Setup} button has been included : Simply select which line in which redshift range to look for and the number of channels to use, {\sf Setup} automatically adjusts {\sf Minimum Frequency} and {\sf Frequency Resolution} for you.

The user also has the possibility of putting in the effects of a Gaussian beam, via the {\sf Beam Major FWHM}, {\sf Beam Minor FWHM} and {\sf Beam Position Angle} entry fields. Please note that a beam smaller than the pixel size specified in the corresponding field has of course no effect. In that case, no convolution is performed.

Gaussian noise can be added to each frequency plane in the output spectral cube. This is done via the dedicated {\sf Noise level} entry field. Currently, the said noise level sets the ratio of the added noise's rms value to the maximal value of the noiseless map.

It is also possible to add the Galactic foreground signal at each frequency form the Global Sky Model described in \ref{sec:gsm}. For this, simply tick the {\sf Global Sky Model} checkbox. Be aware that this is much lower spatial resolution than the SKA Simulated Skies, and so may appear in most cases as just an added constant over the requested skyfield.

The unit of the third axis can be set, using the {\sf Z-axis} menu, to one of frequency, velocity or wavelength. The user should be aware that the "velocity" option only applies to the mapping of a single line. If this option is selected while requesting the mapping of several lines or no line, the script will revert to a frequency axis.
 
The usual convention for the orientation of the right ascension axis is that it increases to the left (i.e. to the east). This can be changed by ticking the {\sf Reverse R.A.} checkbox. 

Finally, to use the mapped field as input to instrument simulators, the user has a possibility to specify the pointing corresponding to the center of the map. This is done using the {\sf Map Center} entry fields at the bottom of the GUI. When the checkbox next to it is off, this pointing is not used and the positions found in the query result are used. Considering that the simulations are centered on (0,0), this may lead to strange behaviour in later stages of the end-to-end simulations, and it is therefore not recommended.

\subsection{Some points specific to $S^3$-SEX}

\subsubsection{Elliptical templates}

In $S^3$-SEX, sources are made up of components that are either ellipses or point sources, and each line in the query results file corresponds to such a component. As mentioned in \ref{sec:mandatoryitems}, the information given for each component includes position, major and minor axes, position angle, continuum fluxes at the reference frequencies, and possibly {\sc Hi} mass. Given a window and a pixel size for the desired map, we can build a template image for each single component by setting to one the values of pixels whose centers fall within the ellipse, and to zero all other pixel values.

\begin{figure}
\centering
\includegraphics[width=8cm]{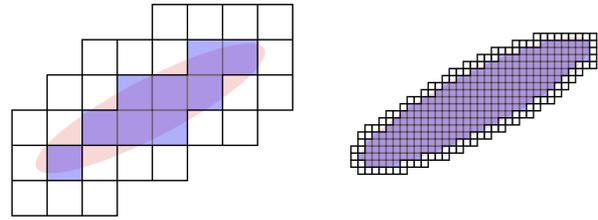}
\caption{Dealing with image accuracy for few-pixel components. {\it Left :} without regridding ; {\it Right :} with regridding.}
\label{fig:accuracy}
\end{figure}

This approach is fine if the component's size is much smaller or much larger than a pixel. In the former case only one pixel is non-zero, and in the latter case, the component's border in the template image resembles the desired ellipse quite closely. A problem arises with components whose sizes are a few pixels across, as can be seen on the left side of Fig.~\ref{fig:accuracy}.

To correct for this, the grid is temporarily refined, as on the right side of Fig.~\ref{fig:accuracy}, with a minimum ratio of semi-major axis to refined pixel size given by the value of the {\sf Image accuracy} entry field on the GUI. Then each coarse pixel in the original grid is given a value proportional to the total number of non-zero refined pixels that belong to this coarse pixel.  

In all cases, the template image is then normalized so that the integrated flux over the component is 1 Jy, and convolved by a flux-conserving beam.

\subsubsection{Continuum and {\sc Hi} fluxes}

Simultaneously, the continuum flux at the desired frequency - or frequencies - is computed from a log-polynomial interpolation based on the reference fluxes retrieved from the query results. 
As for the total {\sc Hi} flux, it has to be derived from the {\sc Hi} gas mass and distance using 
$$
\int \frac{S\mathrm{d}v}{1~\mathrm{Jy.km.s}^{-1}}=4.24\times 10^{-6} \left(\frac{M_{HI}}{M_\odot}\right)\left(\frac{D}{\mathrm{Mpc}}\right)^{-2},
$$
and a synthetic double-peaked profile, is built randomly using some prescriptions based on galaxy type\footnote{For more detail, please refer to the user's guide.}. The {\sc Hi} flux in each frequency channel is computed by integration of this synthetic profile over the channel's width. The total (continuum+{\sc Hi}) flux is then used to multiply the template image at each of the desired frequencies, to obtain a spectral data cube which is finally pasted onto the overall cube.
This means that the emission profile is painted uniformly over the galaxy's disc. This approach is sufficient at low spatial resolution.




\subsubsection{Polarized continuum emission}

By default, only total intensity continuum emission is mapped, but in the {\sf I,Q,U,V} menu, the user may specify that linearly polarized emission should be mapped as well. This emission is computed from the source's total intensity and observing frequencies by routines included in the $S^3$-Tools and which were kindly provided by J\"orn Geisb\"usch (Cavendish Laboratory, Cambridge University). In that case, the output cube has a fourth dimension, with separate Stokes I, Q and U hyperplanes. In the algorithm, for each source, the same template built for total intensity is used for Stokes Q and U, it is simply multiplied by the appropriate fluxes.

\subsubsection{Source types}

There are 5 different source types in the $S^3$-SEX simulation (Radio-quiet AGN, FRI and FRII radio-loud sources, quiescent star-forming galaxies and starbursting galaxies) which are distinguished via their {\sf sftype} and {\sf agntype} attributes. The user can choose to map only selected types by using the {\sf Source Types} pull-down menu. Obviously, only those source types which were queried can be mapped, and the {\sf sftype} and {\sf agntype} must have been included in the query for the selective mapping to be possible. On a related note, GPS sources are a subclass of the FRI and FRII source types (see \cite{wilman08}). The user may specify whether these should be included or excluded, using the {\sf GPS} menu.

\subsection{Some points specific to $S^3$-SAX}

The {\sc Hi} and CO emission maps from $S^3$-SAX data require, as mentioned in \ref{sec:templates}, a number of position-position-velocity (PPV) templates. For {\sc Hi}, the user has the choice between templates from D. Obreschkow ({\sf Oxford} option in the {\sf HI Data} menu) or templates from R. Boomsma ({\sf Kapteyn} option). For CO, only templates by D. Obreschkow are available. They are included in the archive along with the {\sc Hi} templates.

Ten $J\to J-1$ rotational transitions of CO ($J$ going from 1 to 10) are available in the database and can be queried. The queried lines are active in the {\sf Transitions} menu, and by default they are selected for mapping. 

In all cases, for a given galaxy, the relevant template is retrieved based on the values of {\sf hubbletype}, {\sf rmolc}, {\sf himajoraxis\_10max}, {\sf h2majoraxis\_10max}, {\sf diskinclination}, {\sf cowidth50} and {\sf hiwidth50}, then submitted to necessary up- or down-scaling in the spatial and velocity dimensions, rotation and normalization to the integrated line flux. The template is eventually pasted onto the output cube.

\section{Upcoming extensions}

The remaining items on the agenda for the $S^3$-Tools are :

$\bullet$ Inclusion of the epoch of reionization (EoR) signals produced independently by M. Santos (IST Lisbon) and B. Semelin (Observatoire de Paris / LERMA). This will be very similar to the way the GSM is included.

$\bullet$ Inclusion of polarized emission in $S^3$-SAX via templates produced by R. Beck and T. Arshakian (MPIfR)

$\bullet$ The mapping scripts currently deal with one source at a time, building or finding a template, transforming it and pasting it onto the master cube. Obviously, this could be parallelized.

$\bullet$  An older version of the $S^3$-Tools included annotation files for the KARMA visualisation software. This should eventually be re-included.

$\bullet$  The mapping scripts should be installed on the $S^3$ web server and a front-end form similar to the GUI developed for the website. This would allow users to request maps from the server without having to download and install the $S^3$-Tools.

$\bullet$  A better treatment of {\sc Hi} in $S^3$-SEX is possible, for instance by using {\sc Hi} surveys such as HIPASS to derive synthetic line profiles consistent with observations.


\begin{acknowledgements}
This work was supported by the
   European Commission Framework Program 6, Project SKADS, Square
   Kilometre Array Design Studies (SKADS), contract no 011938.
\end{acknowledgements}

\end{document}